\documentclass[aps,prl,twocolumn,amsmath,amssymb,floatfix,showpacs]{revtex4}
\usepackage{epsfig,psfrag,subfigure}
\usepackage{graphicx,psfrag,subfigure}
\usepackage{color}
\newcommand\beq{\begin{equation}}
\newcommand\eeq{\end{equation}}
\newcommand\bea{\begin{eqnarray}}
\newcommand\eea{\end{eqnarray}}
\newcommand\non{\nonumber}

\begin{document}
\title{Double-gap superconducting proximity effect in nanotubes}
\author{Karyn Le Hur$^1$}
\author{Smitha Vishveshwara$^2$}
\author{Cristina Bena$^{3}$}
\affiliation{$^1$ Department of Physics, Yale University, New Haven, CT 06520, USA}
\affiliation{$^2$Department of Physics, University of Illinois at Urbana-Champaign, 1110 W.
Green St, Urbana, IL 61801, USA}
\affiliation{$^3$ Service de Physique Th\' eorique, CEA/Saclay,
Orme des Merisiers, 91190 Gif-sur-Yvette CEDEX}

\date{\today}

\begin{abstract}
We theoretically explore the possibility of a superconducting proximity effect
in single-walled metallic carbon nanotubes due to  the presence of a
superconducting substrate. An unconventional double-gap situation
can arise in the two bands for nanotubes of large radius wherein the
tunneling is (almost) symmetric in the two sublattices. In such a case, a
proximity effect can take place in the symmetric band below a critical
experimentally-accessible Coulomb interaction strength in the nanotube.
Furthermore, due to interactions in the nanotube, the appearance of a BCS gap
in this band stabilizes superconductivity in the other band at lower
temperatures. We also discuss the scenario of highly asymmetric tunneling and
show that this case too supports double-gap superconductivity.
\end{abstract}

\maketitle

\begin{figure}[t]
\includegraphics[width=2.2in,height=1.4in]{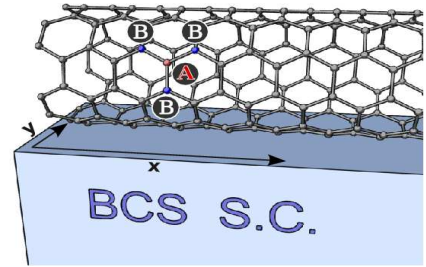}
\caption[]{(color online) A metallic armchair SWMNT, comprising an underlying
graphene structure having sublattices A and B, deposited on an s-wave paired
BCS superconductor. The SWMNT is electron tunnel coupled to the substrate.}
\label{fig:freeNT1}
\end{figure}

Graphene based materials, due to their unique two band structure
\cite{Katsnelson,McEuen}, have recently commanded an explosion of theoretical
and experimental investigations in diverse issues such as Kondo effects and
quantum Hall systems of high SU(4) symmetry \cite{Kondonanotube,QHE}, weak
localization of Dirac fermions \cite{Geim}, Luttinger liquid effects
 \cite{Kane,CS,Bockrath,Yao}, and Coulomb blockade
\cite{Bockrath} in nanotubes. In this Letter, we theoretically investigate the
possibility of a proximity-induced effect in single-walled metallic (carbon)
nanotubes (SWMNTs) due to the presence of a superconducting substrate with
s-wave pairing (see Fig. \ref{fig:freeNT1}) and show that in this geometry the
two-band structure of the SWMNT allows for the appearance of two
superconducting gaps. In the case of nearly symmetric tunneling in the two
sublattices, which may be realized for metallic nanotubes with quite large
radius, we predict two gaps of different origins -  one due to the
superconducting substrate and the other due to electron-electron interactions
inside the SWMNT. For very asymmetric tunneling in the two sublattices, the two
gaps emerge due to the proximity of the substrate and they become identical
when only one sublattice is sensitive to the substrate.

Superconductivity in nanotubes has presented several puzzles. Experimental
observations include a very strong proximity effect in suspended SWMNTs
\cite{Bouchiat}, intrinsic superconductivity in ultrathin nanotubes embedded in zeolite matrix
\cite{Tang}, and BCS type behavior in bundles \cite{Kociak}. It has been predicted that 
superconductivity in an
isolated SWMNT would only be manifest at experimentally inaccessible
temperatures and for screened Coulomb interactions \cite{Egger}. However, 
the phonon exchange might be responsible for some attractive interactions
in the SWMNT \cite{Gonsales,Martino}.  Alternatively, the
theoretical question arises as to whether external factors can also stabilize
superconductivity in SWMNTs, and if so, as to the nature of this
superconducting behavior. In the case of a point contact with a superconductor,
the proximity effect necessarily requires attractive interactions to be
stabilized  \cite{Maslov,Caux}, which then favors a single-gap scenario. Below,
we consider the case of bulk contact with a superconductor and Coulomb
interactions which are unscreened, {\it i.e.}, phonon exchange is not relevant.
We show that a proximity effect is stabilized, and a double
superconducting gap feature appears as a generic unconventional phenomenon.

\begin{figure}[t]
\includegraphics[width=2.2in,height=1.4in]{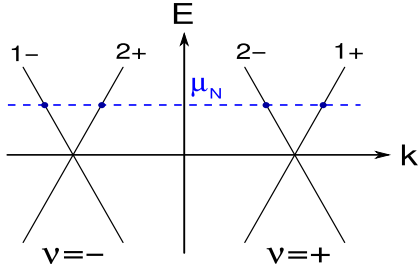}
\caption[]{(color online) Low-energy band structure of a SWMNT. Fermi points are labelled by $\nu=\pm$ and sublattices $A$ and $B$ combine to build two bands with right $(+)$ and left movers $(-)$. Here, $k$ represents the momentum associated with the $x$ direction and $\mu_N$ corresponds to the
Fermi energy.
}
\label{fig:freeNT2}
\end{figure}

As our starting point, we focus on a long, metallic $(N,N)$ armchair nanotube
\cite{Leon}, which is known to be an ideal one-dimensional conductor
\cite{Bockrath,Yao,Devoret}. As shown in Fig.
\ref{fig:freeNT2}, electronic low-energy excitations of the tube consist of two
linearly dispersing bands (labelled by $i=1/2$ and associated pseudo-spin Pauli
matrices $\vec{\eta}$). These bands are symmetric and antisymmetric
combinations of the two sublattices of the nanotube shown in Fig.
\ref{fig:freeNT1}. In the effectively infinite system, each band has an
associated right and left mover (labelled by $r=+/-$, respectively, and
pseudo-spin Pauli matrices $\vec{\tau}$). Each of the modes also carries spin
(labelled by $\alpha=\ \uparrow/\downarrow$ and Pauli matrices $\vec{\sigma}$),
thus constituting eight degrees of freedom.
The Hamiltonian governing these
modes reads
\bea H_0 = iv_F\sum_{ir\alpha}r\int dx\
\psi^{\dagger}_{ir\alpha}\partial_x\psi^{}_{ir\alpha}; \label{eq:ham0}
 \eea
$v_F\approx 8\times 10^5\hbox{m/sec}$ is the Fermi velocity of each of the
modes and $\psi^{\dagger}_{ir\alpha}$ is the operator for creating an electron
of band index `$i$', chirality `$r$', and spin `$\alpha$'. As shown in Fig. 1, the $x$ axis points along the tube direction.

Below, we investigate the effects of external influences such as substrates on these
low-lying SWMNT modes. As the most relevant couplings are expected to be
quadratic in the fermion operators, we first consider their influence before
considering interaction effects in the SWMNT.

As a realistic situation, we explore the case of a nanotube in bulk contact
with a conventional BCS singlet-paired superconducting substrate via electron
tunneling. The simplest such coupling is described by
\bea H_{tun}=\int dx
\left[\chi^{\dagger}_{\alpha}(x,0)(t_A c_{A\alpha}(x)+ t_B
c_{B\alpha}(x))+\hbox{H.c.}\right], \label{eq:hamtunn} \eea
 where an implicit sum on spin $\alpha$ is assumed. Electronic
 substrate degrees of freedom are described by the creation operator
$\chi^{\dagger}_{\alpha}(x,y)$; for simplicity, in Eq. (2), we have assumed a
quasi two-dimensional substrate and the $y$ direction is shown in Fig. 1. The
nanotube degrees of freedom on sublattices $A$ and $B$ are related to the
aforementioned low-lying modes via
$c_{A/B\alpha}^{\nu=+}=(\pm\psi_{2-\alpha}+\psi_{1+\alpha})/\sqrt{2}$ and
$c_{A/B\alpha}^{\nu=-}=(\pm\psi_{2+\alpha}+\psi_{1-\alpha})/\sqrt{2}$. The
matrix elements $t_{A/B}$ related to the electron tunneling strengths into the
two sublattices depend upon the overlap of wavefunctions between the substrate
and nanotube degrees of freedom. Focusing on an armchair SWMNT allows us (in principle) to 
neglect small inhomogeneities in the tunneling amplitudes, {\it i.e.}, $t_A$ and $t_B$ are assumed to be
$x$-independent. 

An effective description of the nanotube under the influence of the substrate
can be obtained by integrating out the substrate degrees of freedom
$\chi_{\alpha}(x,y)$. Such an integration is straightforward given that the
substrate can be described by the standard BCS form \cite{BCS}. For
singlet-paired superconductivity, which is the case of interest below, the main
induced couplings stem from Andreev reflections, resulting in the Hamiltonian
\bea H_{ind} = - \int dx\  \sum_{indices}
(h_i\psi_{ir\alpha}^{\dagger}i\sigma^y_{\alpha\beta}\tau^x_{rw}\psi_{iw\beta}^{\dagger}
\\ \non +
 h_3\psi_{ir\alpha}^{\dagger}i\sigma^y_{\alpha\beta}\eta^x_{ij}\psi_{jr\beta}^{\dagger} + \hbox{H.c.}).
 \eea
 It should be noted that the Andreev term $h_3$, which couples the two bands,
  can only be relevant if the Fermi energy $\mu_N$ of the SWMNT
lies at the Dirac points; otherwise, this process does not conserve momentum
and therefore becomes irrelevant at long wavelengths.  So, we will ignore it
considering the (general) case where $\mu_N$ does not lie at the Dirac points \cite{notenew}.
The bare couplings $h_{1/2}$ take the general  form, $h_{1/2} = h_0(1\pm\sin 2\theta)$.
For a quasi two-dimensional substrate, the coefficient $h_0$ is proportional to $(t_A^2+t_B^2)$ and is inversely proportional to $\sqrt{\Delta}$ where $\Delta$ is the superconducting gap of the substrate. In general the angle $\theta$ obeys $0\leq \theta=\tan^{-1}(t_B/t_A)\leq \pi/4$ \cite{notetheta}. However $\theta\rightarrow \pi/4$ (symmetric tunneling) should hold for large radius armchair tubes (see Fig.\ref{fig:freeNT1}), substrates whose underlying lattices show significant
mismatch with the nanotube lattice, and for specific orientations of the tube
in which the $A$ and $B$ sublattices are equidistant to the substrate. The situation
$\theta\rightarrow 0$ corresponds to the extreme and less physical case where tunneling only
involves one sublattice. 

We argue that our procedure is
well-controlled in the weak-coupling regime, {\it i.e.}, assuming that the
couplings $h_i$ are smaller than the superconducting gap $\Delta$, which then
can be used as the ultraviolet cutoff of the effective theory. It should be
noted that, since an electron in the SWMNT can virtually leak into the
substrate and then tunnel back into the SWMNT, in principle, the electron
propagator of a given band also acquires a finite self-energy. For a 
superconducting substrate, we have checked that this
self-energy evolves smoothly close to the quasiparticle pole. This effect is
always small compared to that of electron-electron interactions on the electron
self-energy \cite{karyn} and thus can be safely ignored.
Also, for the sake of simplicity, the superfluid phase of
the superconducting substrate is set to zero. 

The above arguments indicate that, depending on the microscopic details
of the tunnel coupling with the substrate, the nanotube can 
develop two BCS-type gaps. In what follows we study the effect of Coulomb
interactions on these gaps along the lines of Refs. \cite{Egger,Kane}. We
also show that interactions have a dramatic effect when $\theta\rightarrow \pi/4$, when the coupling with the SC substrate only gives rise to a gap in the symmetric band. In this context interactions 
reinforce interband Cooper processes consisting of
pair hopping from band 1 to band 2 \cite{Balents,Urs,Karyn}, resulting in a
superconducting gap at lower temperatures in band 2 inspite of the vanishing
value of $h_2$. We already like to emphasize that this scenario is still likely to happen for reasonably small deviations around $\theta=\pi/4$.

To begin with, forward scattering processes, in which electrons stay in the
same branch, produce a long-range interaction which couples the total charge
densities \cite{Kane},
\begin{equation}
H_{int} = e^2 \ln(R_s/R) \int dx\ \rho_{tot}^2(x),
\end{equation}
where $\rho_{tot}=\sum_{ir\alpha} \psi^{\dagger}_{ir\alpha}\psi_{ir\alpha}$,
$R$ being the tube radius, and $R_s$ the screening length of Coulomb
interactions. In general, $R_s$ is long compared to $R$ but short compared to
the length of the tube. In the extensively used Luttinger liquid
description \cite{Giamarchi} of the SWMNT \cite{Egger,Kane}, this term contributes to the
Luttinger parameter $g$ of the total charge ``sector'' as ,
\begin{equation}
g = \left[1 +(8e^2/(v_F\pi\hbar))\ln(R_s/R)\right]^{-1/2}.
\end{equation}
In this description, we find the scaling dimension of the operators $h_{1,2}$ to be
$\delta_h=(3+g^{-1})/4$. We deduce that the Andreev terms $h_{1,2}$ are relevant when
$g$ exceeds the critical value $g_c=0.2$, {\it i.e.}, when $\delta_h<2$,
thereby opening BCS-type gaps below the critical temperatures,
\begin{equation}
T_{c1,2} \approx \Delta \left(\frac{h_{1,2}}{\Delta}\right)^{4/(5-g^{-1})}< \Delta.
\end{equation}
As mentioned above, the superconducting gap $\Delta$ in the BCS-type substrate
is the ultraviolet cutoff of our (effective) theory which justifies that
$T_{c2} < T_{c1} < \Delta$. For free electrons, {\it i.e.}, $g=1$, we recover the critical
temperatures $T_{c1,2}\sim h_{1,2}$. Coulomb interactions in the tube tend to reduce
the superconducting gaps $\Delta_{1,2} \sim T_{c1,2}$ which are
experimentally accessible if $g$ is not too close to $0.2$, {\it i.e.}, for a
screening length $R_s$ which is smaller than 1000\AA.  

Now, we investigate in depth the more unconventional, physically accessible situation $\theta\rightarrow \pi/4$ or $h_2\rightarrow 0$ below the critical temperature $T_{c1}$. In this situation $\Delta_2\rightarrow 0$ and apparently only band 1 is gapped.
First off, the superconducting gap $\Delta_1$ in band 1
affects spectral properties such as the local density of states to tunnel an
electron into a long tube at a given site on the tube from a metallic electrode
or a scanning tunneling microscopy (STM) tip. For example, in the
weak-tunneling regime, the (tunneling) current $I$ should exhibit a prominent
peak at the bias voltage $V=\pm \Delta_1/e$ reflecting the profile of the BCS
density of states \cite{Tinkham}. Whereas a BCS gap develops in band 1, close to and below $T_{c1}$, the band 2
charge sector still obeys a Luttinger theory,
\begin{equation}
H(\theta_{2\rho},\varphi_{2\rho}) = \frac{v_F}{2\pi}\int dx \left[ \frac{1}{g_2^{2}}(\partial_x\theta_{2\rho})^2 + (\partial_x\varphi_{2\rho})^2\right],
\end{equation}
where $\partial_x\theta_{2\rho}$ represents the charge density associated with
band 2 and $\varphi_{2\rho}$ embodies the conjugate superfluid phase. The
Luttinger parameter satisfies $g_2^{-2} = (g^{-2} +1)/2$. 

Band 1 and band 2 are
still coupled through $H_{int}$ and the most relevant coupling is of the form
\begin{equation}
H_{int}^{(1)} =\int dx\ v_{12}^{\rho} \partial_x\theta_{1\rho}\partial_x\theta_{2\rho},
\end{equation}
where $v_{12}^{\rho}=v_F(g^{-2}-1)/2\pi$. Given that the superfluid phase
$\varphi_{1\rho}$ of band 1 is pinned below $T_{c1}$, thus inducing strong
fluctuations in the density operator  $\partial_x\theta_{1\rho}$, the coupling
$v_{12}^{\rho}$ renders the $\theta_{2\rho}$ correlation function with
additional fluctuations. This potentially enhances $g_2$. The coupling
$v_{12}^{\rho}$ is in fact an analytic perturbation which has been thoroughly
analyzed in a different context \cite{Urs}. To evaluate the increase of $g_2$
below $T_{c1}$ we proceed as in Ref. \cite{Urs} which gives $g_2^r \approx g_2\left(1-g_2^4(v_{12}^{\rho})^2\pi^2/(2 v_F^2)\right)^{-1}$ \cite{noteg2} and $g_2^r \approx 2g_2 \approx 2\sqrt{2}g$ assuming that $g\ll 1$.  The tunneling current at low
bias voltage emerging from the gapless band 2 thus obeys the form $dI/dV \sim
V^{\zeta}$, where $\zeta=(g_{2}^r+1/g_{2}^r-2)/4$. The exponent $\zeta$ is
distinguishable from the exponent $(g+g^{-1}-2)/8$ which occurs in the absence
of the substrate, {\it i.e.}, above $T_{c1}$ \cite{Kane}.

The results above hold for a range of temperatures below $T_{c1}$. However, as
shown below, at still lower temperatures, several short-range interactions
become important, in particular backward scattering mechanisms and other
forward scattering processes measuring the difference between intra- and
inter-sublattice interactions. In the absence of a superconducting substrate,
these momentum-conserving scattering vertices are ``marginal'' and only become
important at an exponentially small (unreachable) energy scale \cite{Egger}.
However, in the presence of the substrate, below $T_{c1}$, these terms gain relevance as a result of the off-diagonal long-range order in band 1, {\it i.e.},
\begin{equation}
\label{cond}
{\cal P}_1 = \langle \psi^{\dagger}_{1+\uparrow}\psi^{\dagger}_{1-\downarrow} +\psi^{\dagger}_{1-\uparrow}\psi^{\dagger}_{1+\downarrow}\rangle \neq 0,
\end{equation}
where we estimate ${\cal P}_1 \sim T_{c1}/\hbar v_F$. In analogy with the two-chain Hubbard model \cite{Urs,Balents}, we thus anticipate that interband Cooper type processes will be reinforced below $T_{c1}$. We note that Eq. (3) implies ${\cal P}_1>0$, consistent with the superfluid phase of band 1 coinciding with that of the superconducting substrate below $T_{c1}$.

From Ref. \cite{Egger}, we identify two relevant forward $(f)$ and backward $(b)$ scattering processes respecting ${\cal P}_1 \neq 0$,
\begin{eqnarray}
H_{int}^{(2)} =  \int dx  \sum_{\alpha} (-f\psi^{\dagger}_{1+\alpha}\psi^{\dagger}_{1- \bar{\alpha}} \psi_{2-\alpha} \psi_{2+\bar{\alpha}} \non \\
+  b\psi^{\dagger}_{1+\alpha}\psi^{\dagger}_{1- \bar{\alpha}} \psi_{2-\bar{\alpha}} \psi_{2+\alpha}+\hbox{H.c.}),
\end{eqnarray}
with $\bar{\alpha}=\ \downarrow$ if $\alpha=\ \uparrow$, and viceversa. Assuming the equality $b=f$, this results in the following (exact) Hamiltonian
\begin{equation}
H_{int}^{(2)} = \int dx\  f {\cal P}_1 \left(\psi_{2-\downarrow}\psi_{2+\uparrow} + \psi_{2+\downarrow}\psi_{2-\uparrow} +\hbox{H.c.}\right),
\end{equation}
which characterizes pair hopping mechanisms from one band to the other. Using
the notations of Ref. \cite{Egger}, we get $f {\cal P}_1 \approx \gamma 5.4\pi
T_{c1}/(\sqrt{3} N)$ where $\gamma$ is a constant of the order of $0.1$ and $N$
is the circumference of the tube in units of graphene periodicity.  The factor
$1/N$ appears because these terms stem from short-ranged contributions and the
probability for two electrons to be near each other is of order $1/N$. A small
deviation from $b=f$ does not affect the result because, below $T_{c1}$, one
obtains $\langle \psi^{\dagger}_{1+\uparrow}\psi^{\dagger}_{1-\downarrow}
-\psi^{\dagger}_{1-\uparrow}\psi^{\dagger}_{1+\downarrow}\rangle =0$ (as a
result of ${\cal P}_1 \neq 0$). The Cooper pair (Higgs) field associated with
band 1 thus induces a superconducting gap in band 2 and the associated critical
temperature $T_{c3}$ is defined by,
\begin{equation}
T_{c3} \approx T_{c1}\left(\frac{f{\cal P}_1}{T_{c1}}\right)^{2/(3-1/g_{2}^r)} < T_{c1}.
\end{equation}

For $\theta\rightarrow \pi/4$, the emergent superconducting gap in band 2,
$\Delta_3\sim T_{c3}$, is also accessible experimentally;  for a $(10,10)$
armchair nanotube we find $f{\cal P}_1/T_{c1} \sim 0.1$. 
It should be noted that $H_{int}^{(2)}$ implies that the
superconducting order parameters of the two bands exhibit a relative negative
sign which is consistent with repulsive interactions $(f>0)$ and two-band
Hubbard ladders \cite{Balents,Urs,Giamarchi}.

Thus, the band structure of the SWMNT offers two possible mechanisms for a double-gap
superconducting proximity effect. For the extreme asymmetric tunneling case
$\theta\rightarrow 0$, the two bands equally couple to the substrate and each
exhibits a proximity-induced gap. For completely symmetric tunneling
$\theta\rightarrow \pi/4$, even though only the symmetric band becomes affected
by the substrate, interactions in the SWMNT still open a superconducting gap in
the antisymmetric band. We conjecture that the exotic double-gap scenario
occurring at $\theta\rightarrow \pi/4$ can be realized for armchair nanotubes
of large radius and substrates whose underlying lattices show significant
mismatch with the nanotube lattice. This situation is likely to happen as long as $T_{c3}>T_{c2}$.
The two proximity effects associated with
band 2 ``compete'', {\it i.e.}, they give a different sign to the
superconducting order parameter of band 2 and thus result in a quantum phase
transition occurring at the value of $\theta$ for which $T_{c2} = T_{c3}$; at
this special point the band 2 remains gapless until zero temperature. For
temperatures smaller than $T_{c2}$ $(T_{c3})$, the tunneling current exhibits a
complete gap at low bias voltages and a second peak at $V=\pm \Delta_2/e$
($V=\pm \Delta_3/e$).

In conclusion, superconducting substrates stabilize superconductivity in SWMNTs
and allow for the existence of a double-superconducting gap. A double-proximity effect in nanotubes is yet to be ascertained by experiment. Suggestively, several gaps
have been observed in few-layer-graphene coupled to superconducting leads
\cite{Bouchiat2}. Extensions of this work include the proximity effect
between graphene and a (superconducting) substrate, and the study of possible asymmetry in the 
coupling with the substrate for graphene-based materials \cite{graphene}.

We thank L. Balents, H. Bouchiat, C. Dekker, M. P. A. Fisher, T. Giamarchi, and N. Mason for discussions. K. L. H. and S. V. are grateful to the Aspen
Center for Physics where this work has been developed. C. B. acknowledges the
support of a Marie Curie Action under the Seventh Framework Programme, and S. V. the support of NSF
DMR 0605813.



\begin{thebibliography}{99}
\bibitem{notetheta}
The value of $\theta$ and its 
dependence on the radius and the chirality of a tube, as well as on the nature of the substrate are 
important questions, which are now also being brought to light by the debates concerning graphene on a SiC substrate \cite{graphene}.
\end{thebibliography}
\end{document}